\documentclass[prd,twocolumn,aps,superscriptaddress]{revtex4-1}
\usepackage{color,comment}
\usepackage{hyperref}
\usepackage{graphicx}
\usepackage{amsmath}
\usepackage{slashed}

\usepackage{amssymb}
\begin{document}

%


\newcommand{\Gb}{\pazocal{G}}
\newcommand{\be}{\begin{equation}}
\newcommand{\ee}{\end{equation}\noindent}
\newcommand{\bear}{\begin{eqnarray}}
\newcommand{\ear}{\end{eqnarray}\noindent}
\newcommand{\no}{\noindent}
\newcommand{\non}{\nonumber\\}
\newcommand{\red}{\color{red}}
\def\veps#1{\varepsilon_{#1}}
\def\ddel{{}^\bullet\! \Delta}
\def\deld{\Delta^{\hskip -.5mm \bullet}}
\def\dddel{{}^{\bullet \bullet} \! \Delta}
\def\ddeld{{}^{\bullet}\! \Delta^{\hskip -.5mm \bullet}}
\def\deldd{\Delta^{\hskip -.5mm \bullet \bullet}}
\def\epsk#1#2{\varepsilon_{#1}\cdot k_{#2}}
\def\epseps#1#2{\varepsilon_{#1}\cdot\varepsilon_{#2}}
\def\eq#1{{eq. (\ref{#1})}}
\def\eqs#1#2{{eqs. (\ref{#1}) -- (\ref{#2})}}
\def\t#1{\tau_1}
\def\mn{{\mu\nu}}
\def\rs{{\rho\sigma}}
\newcommand{\Det}{{\rm Det}}
\def\Tr{{\rm Tr}\,}
\def\tr{{\rm tr}\,}
\def\sumij{\sum_{i<j}}
\def\e{\,{\rm e}}
\def\eps{\varepsilon}
\def\eps{\varepsilon}
\def\bk{{\bf k}}
\def\bp{{\bf p}}
\def\bb{{\bf B}}
\def\bq{{\bf q}}

\def\bddel{{}^\bullet\! {\underline\Delta}}
\def\bdeld{{\underline\Delta}^{\hskip -.5mm \bullet}}
\def\bdddel{{}^{\bullet \bullet} \! {\underline\Delta}}
\def\bddeld{{}^{\bullet}\! {\underline\Delta}^{\hskip -.5mm \bullet}}
\def\bdeldd{{\underline\Delta}^{\hskip -.5mm \bullet \bullet}}
\def\f{\mathcal{F}}


\def\fr#1#2{\frac{#1}{#2}}
\def\half{\frac{1}{2}}
\def\freeexp{{\rm e}^{-\int_0^Td\tau {1\over 4}\dot x^2}}
\def\kinb{{1\over 4}\dot x^2}
\def\kinf{{1\over 2}\psi\dot\psi}
\def\expk{{\rm exp}\biggl[\,\sum_{i<j=1}^4 G_{Bij}p_i\cdot p_j\biggr]}
\def\expp{{\rm exp}\biggl[\,\sum_{i<j=1}^4 G_{Bij}p_i\cdot p_j\biggr]}
\def\expshort{{\e}^{\half G_{Bij}p_i\cdot p_j}}
\def\expabb{{\e}^{(\cdot )}}
\def\epseps#1#2{\varepsilon_{#1}\cdot \varepsilon_{#2}}
\def\epsk#1#2{\varepsilon_{#1}\cdot k_{#2}}
\def\epsr#1#2{r_{#2}\cdot\varepsilon_{#1}}
\def\rk#1#2{r_{#1}\cdot p_{#2}}
\def\kk#1#2{k_{#1}\cdot k_{#2}}
\def\G#1#2{G_{B#1#2}}
\def\Gp#1#2{{\dot G_{B#1#2}}}
\def\GF#1#2{G_{F#1#2}}
\def\Dab{{(x_a-x_b)}}
\def\Dsq{{({(x_a-x_b)}^2)}}
\def\PITD{{(4\pi T)}^{-{D\over 2}}}
\def\4piTD{{(4\pi T)}^{-{D\over 2}}}
\def\4piT4{{(4\pi T)}^{-2}}
\def\TintmD{{\dps\int_{0}^{\infty}}{dT\over T}\,e^{-m^2T}
    {(4\pi T)}^{-{D\over 2}}}
\def\Tintm4{{\dps\int_{0}^{\infty}}{dT\over T}\,e^{-m^2T}
    {(4\pi T)}^{-2}}
\def\Tintm{{\dps\int_{0}^{\infty}}{dT\over T}\,e^{-m^2T}}
\def\Tint{{\dps\int_{0}^{\infty}}{dT\over T}}
\def\np{n_{+}}
\def\nm{n_{-}}
\def\Np{N_{+}}
\def\Nm{N_{-}}
\def\ed{e^{(\cdot)}}
\def\t#1{\tau_{#1}}
\def\et#1#2{e^{({#1}\rightarrow{#2})}}
\def\ett#1#2#3#4{e^{({#1}\rightarrow{#2}, #3\rightarrow#4)}}
\newcommand{\slG}{{{\dot G}\!\!\!\! \raise.15ex\hbox {/}}}
\newcommand{\Gd}{{\dot G}}
\newcommand{\Gund}{{\underline{\dot G}}}
\newcommand{\Gdd}{{\ddot G}}
\def\GBd12{{\dot G}_{B12}}
\def\Dx{\dps\int{\cal D}x}
\def\Dy{\dps\int{\cal D}y}
\def\Dpsi{\dps\int{\cal D}\psi}
\def\dint#1{\int\!\!\!\!\!\int\limits_{\!\!#1}}
\def\ddtau{{d\over d\tau}}
\def\ie{\hbox{$\rm style{\int_1}$}}
\def\iz{\hbox{$\rm style{\int_2}$}}
\def\id{\hbox{$\rm style{\int_3}$}}
\def\ldop{\hbox{$\lbrace\mskip -4.5mu\mid$}}
\def\rdop{\hbox{$\mid\mskip -4.3mu\rbrace$}}
\def\bdel{{\underline\Delta}}
%
\newcommand{\1}{{\'\i}}
\def\dps{\displaystyle}
\def\sy{\scriptscriptstyle}
\def\sy{\scriptscriptstyle}

\def\del{\partial}
\def\deli{\partial_{\kappa}}
\def\delj{\partial_{\lambda}}
\def\delk{\partial_{\mu}}
\def\delij{\partial_{\kappa\lambda}}
\def\delik{\partial_{\kappa\mu}}
\def\deljk{\partial_{\lambda\mu}}
\def\delki{\partial_{\mu\kappa}}
\def\delkl{\partial_{\mu\nu}}
\def\delijk{\partial_{\kappa\lambda\mu}}
\def\deljkl{\partial_{\lambda\mu\nu}}
\def\delikl{\partial_{\kappa\mu\nu}}
\def\delijkl{\partial_{\kappa\lambda\mu\nu}}
\def\delijklm{\partial_{\kappa\lambda\mu\nu o}}
\def\O(#1){O($T^#1$)} 
\def\O2{O($T^2$)}
\def\O3{O($T^3$)}
\def\O4{O($T^4)}
\def\O5{O($T^5$)}
\def\dA{\partial^2}
\def\DA{\sqsubset\!\!\!\!\sqsupset}
\def\eins{  1\!{\rm l}  }
\def\a#1{\alpha_{#1}}
\def\b#1{\beta_{#1}}
\def\m#1{\mu_{#1}}
\def\n#1{\nu_{#1}}
\def\m#1{\mu_{#1}}
\def\n#1{\nu_{#1}}
\def\a{\alpha}
\def\b{\beta}
\def\m{\mu}
\def\n{\nu}
\def\s{\sigma}
\def\r{\rho}
\def\e{{\rm e}}
\def\z{\zeta}
\def\vareps{\varepsilon}

\def\gF{\gamma_{\mathcal{F}}}
\def\gG{\gamma_{\mathcal{G}}}
\def\gFF{\gamma_{\mathcal{F}\mathcal{F}}}
\def\gGG{\gamma_{\mathcal{G}\mathcal{G}}}
\def\gFG{\gamma_{\mathcal{F}\mathcal{G}}}

\newcommand{\Vka}{V_{\kappa}}
\newcommand{\Vla}{V_{\lambda}}
\newcommand{\Vmu}{V_{\mu}}
\newcommand{\Vnu}{V_{\nu}}
\newcommand{\Vro}{V_{\rho}}
\newcommand{\Vkala}{V_{\kappa\lambda}}
\newcommand{\Vkamu}{V_{\kappa\mu}}
\newcommand{\Vkanu}{V_{\kappa\nu}}
\newcommand{\Vlamu}{V_{\lambda\mu}}
\newcommand{\Vlanu}{V_{\lambda\nu}}
\newcommand{\Vlaka}{V_{\lambda\kappa}}
\newcommand{\Vmunu}{V_{\mu\nu}}
\newcommand{\Vmuka}{V_{\mu\kappa}}
\newcommand{\Vnuro}{V_{\nu\rho}}
\newcommand{\Vkalamu}{V_{\kappa\lambda\mu}}
\newcommand{\Vkalanu}{V_{\kappa\lambda\nu}}
\newcommand{\Vkalaro}{V_{\kappa\lambda\rho}}
\newcommand{\Vkamunu}{V_{\kappa\mu\nu}}
\newcommand{\Vlamunu}{V_{\lambda\mu\nu}}
\newcommand{\Vmunuro}{V_{\mu\nu\rho}}
\newcommand{\Vkalamunu}{V_{\kappa\lambda\mu\nu}}
\newcommand{\Fkala}{F_{\kappa\lambda}}
\newcommand{\Fkanu}{F_{\kappa\nu}}
\newcommand{\Flaka}{F_{\lambda\kappa}}
\newcommand{\Flamu}{F_{\lambda\mu}}
\newcommand{\Fmunu}{F_{\mu\nu}}
\newcommand{\Fnumu}{F_{\nu\mu}}
\newcommand{\Fnuka}{F_{\nu\kappa}}
\newcommand{\Fmuka}{F_{\mu\kappa}}
\newcommand{\Fkalamu}{F_{\kappa\lambda\mu}}
\newcommand{\Flamunu}{F_{\lambda\mu\nu}}
\newcommand{\Flanumu}{F_{\lambda\nu\mu}}
\newcommand{\Fkamula}{F_{\kappa\mu\lambda}}
\newcommand{\Fkanumu}{F_{\kappa\nu\mu}}
\newcommand{\Fmulaka}{F_{\mu\lambda\kappa}}
\newcommand{\Fmulanu}{F_{\mu\lambda\nu}}
\newcommand{\Fmunuka}{F_{\mu\nu\kappa}}
\newcommand{\Fkalamunu}{F_{\kappa\lambda\mu\nu}}
\newcommand{\Flakanumu}{F_{\lambda\kappa\nu\mu}}

\newcommand{\tvec}{\vec}
\newcommand{\action}{\mathscr{S}}
\newcommand{\pathdiff}[1]{\!\mathscr{D}#1\,}
\newcommand{\diff}[1]{\!\mathrm{d}#1\,}
\newcommand{\dottau}{\accentset{\boldsymbol\circ}}
\newcommand{\dott}{\accentset{\mbox{\large .}}}
\newcommand{\ddott}{\accentset{\mbox{\large ..}}}
\newcommand{\dd}[2][]{\frac{\mathrm{d} #1}{\mathrm{d} #2}}
\newcommand{\pdd}[2][]{\frac{\partial #1}{\partial #2}}
\newcommand{\atanh}{\operatorname{atanh}}
\newcommand{\sech}{\operatorname{sech}}
\newcommand{\keld}{\tilde{\gamma}}
\renewcommand{\Re}{\operatorname{Re}}
\renewcommand{\Im}{\operatorname{Im}}
\newcommand\numberthis{\addtocounter{equation}{1}\tag{\theequation}}

\newcommand{\ket}[1]{\left|#1\right>}
\newcommand{\bra}[1]{\left<#1\right|}
\newcommand{\braket}[2]{\left<#1|#2\right>}
\newcommand{\nn}{\nonumber\\}
\newcommand{\ul}{\underline}
\newcommand{\fk}[1]{\mbox{\boldmath$\scriptstyle#1$}}
\newcommand{\vau}{\mbox{\boldmath$v$}}
\newcommand{\na}{\mbox{\boldmath$\nabla$}}
\newcommand{\bea}{\begin{eqnarray}}
\newcommand{\ea}{\end{eqnarray}}
\newcommand{\eea}{\end{eqnarray}}
\newcommand{\ord}{\,{\cal O}}
\newcommand{\li}{\,\widehat{\cal L}}
\newcommand{\vc}[1]{\mathbf{#1}}
\newcommand{\sumint}[1]
{\begin{array}{c} \\
{{\textstyle\sum}\hspace{-1.1em}{\displaystyle\int}}\\
{\scriptstyle{#1}}
\end{array}}

\newcommand{\new}[1]{{\color{blue} #1}}

\newcommand{\ralf}[1]{{\color{red} #1}}

\title{Assisted neutrino pair production in  combined external fields}

\author{Naser Ahmadiniaz} 
\affiliation{Helmholtz-Zentrum Dresden-Rossendorf, Bautzner Landstra\ss e 400, 01328 Dresden, Germany} 
\author{Rashid Shaisultanov}
\affiliation{Helmholtz-Zentrum Dresden-Rossendorf, Bautzner Landstra\ss e 400, 01328 Dresden, Germany} 
\affiliation{Extreme Light Infrastructure ERIC, Za Radnic\`{i} 835, 25241 Doln\`{i} B\v{r}e\v{z}any, Czech Republic} 
\author{Ralf Sch\"utzhold}
\affiliation{Helmholtz-Zentrum Dresden-Rossendorf, Bautzner Landstra\ss e 400, 01328 Dresden, Germany} 
\affiliation{Institut f\"ur Theoretische Physik, Technische Universit\"at Dresden, 01062 Dresden, Germany}

\begin{abstract}
Neutrino--antineutrino ($\nu\bar\nu$) pair production is one of the main processes responsible for the energy 
loss of stars. 
Apart from the collision of two ($\gamma\gamma\to\nu\bar\nu$) or three ($\gamma\gamma\gamma\to\nu\bar\nu$) 
real photons, pair creation from a photon and photon collisions in the presence of nuclear Coulomb fields or external magnetic 
fields have been considered previously. Here, we study a pair production of neutrino and antineutrino from a low-energy photon in the presence of a 
combined homogeneous magnetic field and the Coulomb field of a nucleus with charge number $Z$.

%
\end{abstract}

\date{\today}

\maketitle

\section{Introduction}\label{intro}

Due to the large mass of the $W^\pm$ and $Z^0$ vector bosons mediating the weak interaction, the coupling of 
neutrinos to the remaining matter particles is extremely suppressed at low energies. 
As a consequence, the cross sections for neutrino--antineutrino pair creation in photon collisions, 
for example, are very small. 
However, because the stellar plasma is basically opaque for electrons and photons but transparent for neutrinos, 
such processes, albeit rare, are very important for the energy loss of stars. 
Neutrinos can carry away energy from the entire volume of the star, whereas electromagnetic radiation can only 
escape from a small surface layer. 
Thus, for large stars with high temperatures $\ord(10^9~\rm K)$  ($\approx 86$ keV) and densities $\ord(10^5~\rm g/cm^3)$, 
processes involving neutrinos may even dominate electromagnetic radiation from the star \cite{adams-63,bethe-85,yakov-01,janka-17,bhat-15}.

Although neutrinos can also be created in processes involving real electrons or nuclear transitions \cite{Gamow,pont1,mati-62,ngu-63,feyn-58,suda-58,chiu-60,band-68}, 
we focus on neutrino pair production by photons or electromagnetic fields in the following. 
A single photon alone can obviously not emit a neutrino pair due to energy-momentum conservation, 
so the lowest-order process corresponds to the collision of two photons $\gamma\gamma\to\nu\bar\nu$.
Since neutrinos do not carry electric charge and thus do not couple directly to photons, this 
process requires an internal charged particle mediating the interaction. 
Here, we consider electrons or positrons as the lightest charged particles, but other charged 
particles such as muons can be treated in complete analogy. 

\begin{figure}[h]
\centering 
\includegraphics[scale=.5]{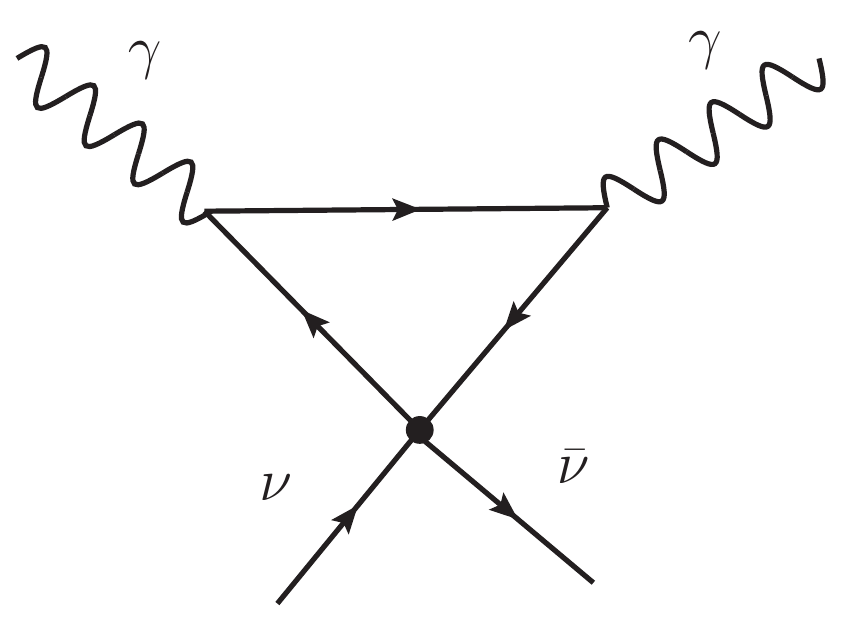}
\caption{Triangle Feynman diagram representing the neutrino antineutrino creation $\gamma\gamma\to\nu\bar\nu$ 
in the four-fermion limit.}
\label{four-fermi}
\end{figure} 

Using the effective low-energy description of the four-fermion interaction, the lowest-order Feynman diagram 
of this process $\gamma\gamma\to\nu\bar\nu$ is depicted in Fig.~\ref{four-fermi}. 
However, it turns out that the associated amplitude is suppressed due to the Landau-Yang theorem \cite{gell,landau,yang,cung}. 
The four-fermion description represents a point interaction (current-current theory) between massless neutrinos 
and thus the two outgoing neutrinos have no relative orbital angular momentum.
Due to the (V-A) coupling structure, their spins must be parallel and thus the final state has total 
angular momentum $J=1$, which cannot couple to two initial (real, i.e., on-shell) photons. 

\begin{figure}
\centering 
\includegraphics[scale=.5]{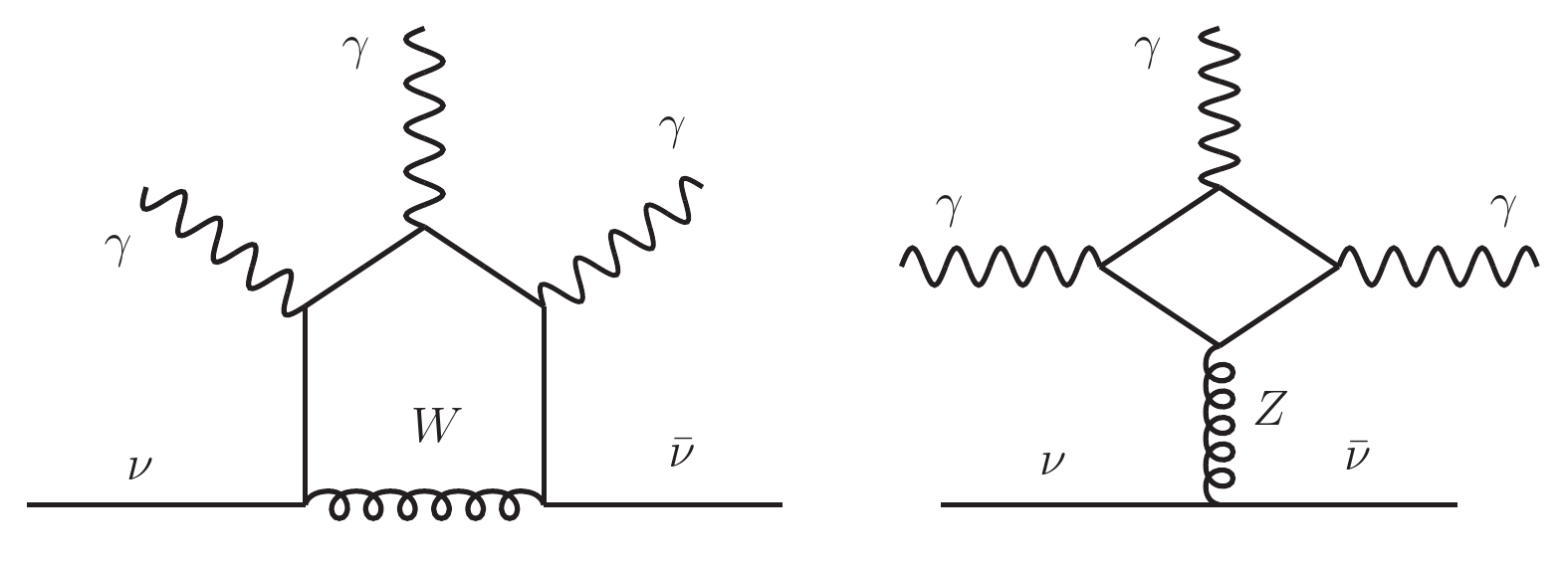}
\caption{Exemplary Feynman diagrams involving neutrino pairs and three photons at one fermion loop order.}
\label{fig1}
\end{figure}

Deviations from the conditions mentioned above, such as taking into account the small neutrino masses \cite{crew-82,dode-91}
or the non-local structure \cite{lev-67,dicrep-93,dicrep-97} of the four-fermion interaction 
(mediated by internal $W^\pm$ or $Z^0$ bosons)
or replacing one of the initial (real) photons by an external field \cite{ros-63,shai-98,ioaraf-97,dicrep-00,tzu-00,shai-00} can lead to a non-zero (albeit small) 
amplitude. 
Another option is a third photon vertex (representing a real photon or an external field), which is the 
case we consider here. 
Exemplary Feynman diagrams are depicted in Fig.~\ref{fig1}, where we have explicitly included the internal 
$W^\pm$ or $Z^0$ bosons instead of using the effective four-fermion interaction as in Fig.~\ref{four-fermi}.

For typical densities, collisions of three real photons can be quite rare events. 
Thus, in order to increase the interaction volume, one may replace one or two initial real photons by external 
electromagnetic fields, which amounts to considering photon-photon collisions or photon decay in this background. 
Examples for such background fields are the Coulomb fields of nuclei or strong magnetic fields, which may also 
exist in stellar environments.  

In the following, we study photon decay into a neutrino--antineutrino pair
in the combined background of a strong magnetic field superimposed by a Coulomb field.
In analogy to a recent study on Coulomb assisted vacuum birefringence \cite{ahma-20},
such a scenario offers advantages in comparison to other cases, such as an 
enlarged interaction volume. 
Furthermore, for photon energies way below the MeV scale, all involved scales 
are subcritical (i.e., also well below MeV) and thus we may use a low-energy 
effective description.


\section{Modified Euler-Heisenberg Lagrangian}\label{Euler-Heisenberg}

To begin, we will provide a concise summary of the fundamental principles. The Euler-Heisenberg Lagrangian outlines a direct connection between low-energy photons and serves as a quantum adjustment to classical Maxwell electromagnetism, building on the pioneering work of Euler and Heisenberg \cite{EH}.  
When the electromagnetic fields represented by the field strength tensor $F_{\mu\nu}$ are slowly varying and remain significantly lower than the Schwinger critical field, which is determined by the electron mass $m_e$, elementary charge $q$, and the related critical magnetic field

\bear
E_{\rm crit}&=&\frac{m_e^2c^3}{\hbar q}\approx 1.3\times10^{18}\,\frac{\rm V}{\rm m}\,,\non
B_{\rm crit}&=&\frac{m_e^2c^2}{\hbar q}\approx 4.4\times10^{9}\,{\rm T}\,,
\ear
then one can examine the low-energy limit of the Euler-Heisenberg Lagrangian. 
In this limit which is of interest in what follows, the one-loop photon-photon interaction is described by the lowest-order in the Euler-Heisenberg Lagrangian \cite{EH,euko-35,sauter-31,karneu-51,dunne-12,4-photon1} given by 
\bear
\label{EHL}
\mathcal{L}^{(1)}&=&\xi\Big[\frac{1}{4}(F_{\mu\nu}F^{\mu\nu})^2+\frac{7}{16}(F_{\mu\nu}\tilde{F}^{\mu\nu})^2\Big]\non
&=&\frac{2\xi}{16}\Big[14\,{\rm tr}(F^4)-5\,({\rm tr}(F^2))^2\Big]\,,
\ear
with the pre-factor 
\bea
\xi=\frac{\hbar q^4}{360\pi^2 m_e^4c^7}=\varepsilon_0\,\frac{\alpha_{\rm QED}}{90\pi E_{\rm crit}^2}=\frac{2\alpha_{\rm QED}^2}{45m_e^4}
\,,
\ea
To examine the range of validity for the low field approximation provided by the Euler-Heisenberg Lagrangian in various physical regimes, see \cite{baier-18,aleks-19}. 
The electromagnetic field strength tensor, denoted by $F_{\mu\nu}$, and its dual, represented by $\tilde{F}^{\mu\nu}=\half \epsilon^{\mu\nu\alpha\beta}F_{\alpha\beta}$, are related to each other.
For recent experimental attempts of light-by-light scattering, see \cite{atlas1,atlas2,atlas3,atlas4}.
From now on, we shall employ natural units with 
\bea
\hbar=c=\varepsilon_0=1\,,
\ea
in order to simplify the expressions. In deriving the last equality in (\ref{EHL}) we have used the following identity which links the field strength tensor with its dual 
\bear
(F\tilde{F})^2=4\,{\rm tr}(F)^4-2({\rm tr}(F^2))^2\,,
\ear
where the trace is taken over Lorentz indices.
As previously noted, any violation of the constraints set by the Landau-Yang theorem results in finite contributions to the scattering amplitude that involves neutrinos. In the absence of external fields and nuclei, the first non-vanishing amplitudes of interest involving two neutrinos are those that have three photons, as presented below
\bear
\gamma\nu\rightarrow\gamma\gamma\nu\label{am1}\,,\\
\gamma\gamma\rightarrow\gamma\nu\bar{\nu}\label{am2}\,,\\
\nu\bar{\nu}\rightarrow\gamma\gamma\gamma\label{am3}\,.\ear

The effective Lagrangian presented below was derived for the photon-neutrino interactions involving five points in \cite{dicrep-93,dicrep-97} and justified in \cite{aba-98}

\bear
\mathcal{L}_{\rm eff}&=&\frac{G_Fa}{\sqrt{32\pi}m_e}\Big(\frac{2\xi^3}{45}\Big)^{\frac{1}{4}} N_{\mu\nu}\non
&&\times\Big\{5F^{\mu\nu}F_{\lambda\rho}F^{\lambda\rho}-14F^{\nu\lambda}F_{\lambda\rho}F^{\rho\mu}\Big\}\,,
\label{ef1}
\ear
where 
\bear
N_{\mu\nu}=\partial_\mu\Big[\bar{\psi}\,\gamma_\nu(1+\gamma_5)\,\psi\Big]-\partial_\nu\Big[\bar{\psi}\,\gamma_\mu(1+\gamma_5)\,\psi\Big]\,,\non
\label{n}
\ear
is the effective coupling for the electron-neutrino part with the fermion-loop via $W$ or $Z$ bosons, 
$a=1-\half(1-4\sin^2\theta_W)$ 
where $\theta_{\rm W}$ is the weak mixing angle or Weinberg angle ($\sin^2\theta_{ W}=1-(m_{ W}/m_{ Z})^2\sim 0.23142$), \cite{glashow-61,lhcb}. The first term in $a$ ($a_{ W}=1$) is for $W$ and the last ($a_{ Z}\sim 0.04$) for the $Z$ boson contribution, 
see Fig.~\ref{fig1} for the corresponding Feynman diagrams. The reduced value of the Fermi constant $G_F$ is expressed as 
$G_F/(\hbar c)^3= 1.1663787(6)\times 10^{-5}~{\rm GeV^{-2}}$ in natural units \cite{chit-07}. 
The effective Lagrangian mentioned earlier is the result of substituting the neutrino current $N_{\mu\nu}$ for one of the external legs in the four-photon amplitude derived from the EH effective action at low energy. In a previous work \cite{shai-98}, the process of producing a neutrino-antineutrino pair was examined in the presence of an external magnetic field, which replaced one of the incoming photons. This external magnetic field was found to enhance the cross section compared to vacuum processes, as discussed earlier. The calculation of the cross section for neutrino pair production under the condition where the magnetic field is perpendicular to the incoming photons and the momenta of the neutrino particles have been integrated out is given by \cite{shai-98}

\bear
\sigma_B\Big\vert_{\gamma\gamma\rightarrow\nu\bar{\nu}}\propto 10^{-51}\Big(\frac{\omega}{m_e}\Big)^6\,\Big(\frac{B}{B_{\rm crit}}\Big)^2\,{\rm cm}^2\,.
\ear
The comparison between the scattering in a constant magnetic field and in vacuum results in a ratio of 
\bear
\frac{\sigma_B(\gamma\gamma\rightarrow \nu\bar{\nu})}{\sigma(\gamma\gamma\rightarrow \nu\bar{\nu})}\propto \Big(\frac{m_W}{m_e}\Big)^4\,\Big(\frac{B}{B_{\rm crit}}\Big)^2\,.
\ear
In this article, we will perform the same calculation for the production of a neutrino-antineutrino pair in a mixed background consisting of a constant magnetic field and Coulomb field. This process involves the decay of an incoming photon into a pair of neutrino and antineutrino in the presence of the mixed background, which can be represented diagrammatically by Fig. \ref{fig1} after replacing two of the photons with those from the external fields. It is worth noting that the presence of the Coulomb field results in an additional enhancement compared to the pure magnetic field case discussed earlier.

The effective Lagrangian is altered by the presence of the mixed magnetic and Coulomb fields. To account for this modification, we begin by replacing the strength tensor in (\ref{ef1}) with a sum of three terms
\bear
f_{\mu\nu}\rightarrow f_{\mu\nu}+F_{\mu\nu}+\f_{\mu\nu}\,.
\ear
Here, $f$, $F$, and $\f$ represent the field strength tensors for the incoming photon, the magnetic field, and the Coulomb field, respectively. To obtain the desired outcome for the process of interest, it is necessary to include multilinear terms involving all three field strength tensors. By doing so, we obtain 
\bear
&&\mathcal{L}_{\rm eff}=\frac{G_Fa}{\sqrt{32\pi} m_e}\,\Big(\frac{2\xi^3}{45}\Big)^{\frac{1}{4}}N_{\mu\nu}\non
&&\times\bigg\{-10\Big[F^{\mu\nu}{\rm tr}(\f f)+\f^{\mu\nu}{\rm tr}(Ff)+f^{\mu\nu}{\rm tr}(\f F)\Big]\non
&&-14\Big[F^{\nu\lambda}(\f_{\lambda\rho}f^{\rho\mu}+f_{\lambda\rho}\f^{\rho\mu})+\f^{\nu\lambda}(F_{\lambda\rho}f^{\rho\mu}+f_{\lambda\rho} F^{\rho\mu})\non
&&\hspace{2cm}+f^{\nu\lambda}(F_{\lambda\rho}\f^{\rho\mu}+\f_{\lambda\rho}F^{\rho\mu})\Big]\bigg\}\,.
\label{effn1}
\ear
Substituting the effective coupling of the electron-neutrino, $N_{\mu\nu}$, back into the calculation, we obtain the following expression for the amplitude
\bear
\mathcal{M}_{BN}=\frac{G_Fa}{\sqrt{8\pi} m_e}\,\Big(\frac{2\xi^3}{45}\Big)^{\frac{1}{4}}\,\bar{u}(p_1)\gamma_\mu\,\mathbb{J}^\mu(1+\gamma_5)\,v(p_2)\,.\non
\label{samp1}
\ear
The neutrino pair spinors are denoted by $\bar{u}$ and $v$. It is important to note that, unlike the previous findings presented in \citep{shai-98}, in this case, the derivative in $N_{\mu\nu}$ acts not only on the quantum field $f_{\alpha\beta}$ but also on the Coulomb field $\f_{\alpha\beta}$. Thus, it is possible to decompose the total current into two distinct contributions
\bear
\mathbb{J}^\mu=\mathbb{J}_f^\mu+\mathbb{J}_\f^\mu\,.
\label{currents}
\ear
We first write down the contribution from $f$:
\bear
\hspace{-0.5cm}\mathbb{J}_f^\mu&=&6i\Big[(F\cdot k)^\mu(\varepsilon\cdot \f\cdot k)+(F\leftrightarrow \f)\Big]\non
&&\hspace{-1cm}+14i\Big\{2\varepsilon^\mu(k\cdot F\cdot\f\cdot k)-k^\mu\Big[(k\cdot\f\cdot F\cdot \varepsilon)+(F\leftrightarrow \f)\Big]\Big\}\,.\non
\label{jf}
\ear
To evaluate $\mathbb{J}_\f$ which corresponds to contribution from the derivative correction of the Coulomb field, we must first consider the vector potential for the Coulomb field, 
its corresponding field strength tensor, and ultimately its Fourier transform. 
The vector potential for the Coulomb field is expressed as follows

\bear
A_\mu=\left(-\frac{Ze}{4\pi r},{\bf 0}\right)
\,.
\ear
To compute $\mathbb{J}_{\mathcal{F}}$ we need \cite{frahm}
\bear
\partial_i\f_{j0}
=
-\frac{Ze}{4\pi}\partial_i\partial_j\frac{1}{r}
\,.
\label{dF}
\ear
Its Fourier transform reads (see, e.g., \cite{berlifpit,breit-29,adkins-13}) 
\bear 
\int d^3r\e^{i\Delta{\bf q}\cdot {\bf r}}\,\partial_i\f_{j0}
=
Ze\frac{(\Delta{\bf q})_i (\Delta{\bf q})_j}{\vert\Delta\bf q\vert^2}\,,
\label{deriv}
\ear 
where $\Delta{\bf q}$ represents the momentum that is exchanged during the current process of assisted 
neutrino pair production.
For arbitrary momentum transfer $\Delta{\bf q}$, this Fourier transform is bounded from above by the 
constant $Q=Ze$.
In contrast to Eq. (\ref{deriv}) which is the Fourier transform of the derivative of the Coulomb field, Eq.~(\ref{ft-coulomb}) shows that the corresponding Fourier transform of the Coulomb field 
itself, which determines $\mathbb{J}_f$, can become arbitrarily large for small $\Delta{\bf q}$,
i.e., in forward direction. 
This growth for small $\Delta{\bf q}$ corresponds to the large effective interaction volume. 
In this limit, the current $\mathbb{J}_{\f}$ is less important than $\mathbb{J}_f$ and thus we 
neglect it in the following.

The square of the amplitude in (\ref{samp1}) results in:
\bear
\vert \mathcal{M}_{BN}\vert^2
&=&\frac{8G_F^2a^2\alpha_{\rm QED}^3}{\pi(180)^2}\frac{1}{m_e^8}\, \Big(\bar{u}(p_1)\gamma^\mu\mathbb{J}_{f\mu}(1-\gamma^5)v(p_2)\Big)\non
&&\times \Big(\bar{v}(p_2)(1+\gamma^5)(\mathbb{J}_{f\mu})^*\gamma^\mu u(p_1)\Big)\,.
\label{samp2}
\ear 
To compute the differential cross section, we sum up the final states of the neutrino pair spins and average the initial polarization states of the incoming photon ($\varepsilon$) 
\bear
&&\half \sum_{s,s';\,\varepsilon}\vert \mathcal{M}_{BN}\vert^2=\frac{8G_F^2a^2\alpha_{\rm QED}^3}{2\pi(180)^2}\frac{1}{m_e^8}\non
&&\times{\rm tr}\Big[(\slashed{p}_1+m)(\gamma\cdot\mathbb{J}_{f})(1-\gamma^5)(\slashed{p}_2-m)(1+\gamma^5)(\mathbb{J}_{f}^*\cdot\gamma)\Big]\,.\non
\ear 
There trace reults in three contributions 
\bear
{\rm tr}\Big[\cdots\Big]=8\Big[p_1\cdot\mathbb{J}_f\,p_2\cdot\mathbb{J}_f^*-p_1\cdot p_2\,\mathbb{J}_f\cdot\mathbb{J}_f^*+p_1\cdot\mathbb{J}_f^*\,p_2\cdot\mathbb{J}_f\Big]\,.\non
\ear
Plugging all back into (\ref{samp2}) and averaging over the incoming on-shell photon polarization we arrive at 

\begin{widetext}
\bear
\half\sum_{s,s';\,\varepsilon}\vert \mathcal{M}_{BN}\vert^2&=&\frac{32G_F^2a^2\alpha_{\rm QED}^3}{\pi(180)^2}\frac{1}{m_e^8}
\bigg\{p_1\cdot p_2\Big[72(k\cdot F^2\cdot k)(k\cdot\f^2\cdot k)+968\vert k\cdot F\cdot\f\cdot k\vert^2\Big]\non
&&\hspace{0cm}+72\Big[(p_1\cdot F\cdot k)(p_2\cdot F\cdot k)\,(k\cdot\f^2\cdot k)+(F\leftrightarrow\f)+(p_1\cdot F\cdot k)(p_2\cdot \f\cdot k)(k\cdot F\cdot \f\cdot k)+(F\leftrightarrow\f)\Big]\non
&&-392\Big[-2p_2\cdot k(k\cdot F\cdot\f\cdot k)(p_1\cdot F\cdot\f\cdot k+p_1\cdot \f\cdot F\cdot k)-2p_1\cdot k(k\cdot F\cdot \f\cdot k)(p_2\cdot\f\cdot F\cdot k+p_2\cdot F\cdot \f\cdot k)\non
&&\hspace{2cm}+p_1\cdot k\,p_2\cdot k\Big((k\cdot \f\cdot F^2\cdot \f\cdot k)+(F\leftrightarrow \f)\Big)\Big]\non
&&-336\Big[(p_1\cdot F\cdot k)(p_2\cdot \f\cdot k)(k\cdot F\cdot \f\cdot k)+(F\leftrightarrow \f)+(p_1\cdot \f\cdot k)(p_2\cdot F\cdot k)(k\cdot F\cdot \f\cdot k)+(F\leftrightarrow \f)\Big]\non
&&+14\times 6\times 2(p_1\cdot k)\Big[(p_2\cdot F\cdot k)(k\cdot F\cdot\f^2\cdot k)+(F\leftrightarrow\f)\Big]\non
&&+14\times 6\times2(p_2\cdot k)\Big[(p_1\cdot \f\cdot k)(k\cdot\f\cdot F^2\cdot k)+(F\leftrightarrow\f)\Big]\bigg\}\,.\non
\label{samp3}
\ear
\end{widetext}

Assuming that the external magnetic field is constant $({\bf B})$ and the nuclear Coulomb field is given by ${\bf E}={\bf e}_r Q/(4\pi {\bf r}^2)$ with charge $Q=Zq$, where ${\bf e}_r$ is the unit vector in the radial direction, we can calculate its Fourier transform
\bear
\int d^3r\,\e^{i\Delta{\bf q}\cdot {\bf r}}\,{\bf E}=\int d^3r\,\e^{i\Delta{\bf q}\cdot {\bf r}}\,\frac{Q\,{\bf e}_r}{4\pi {\bf r}^2}=iQ\frac{\Delta{\bf q}}{\vert\Delta{\bf q}\vert^2}\,.
\label{ft-coulomb}
\ear
The above squared amplitude can be expressed in the following manner, based on the electric field of the nuclei and the constant magnetic field
\begin{widetext}
\bear
\half \sum_{s,s';\,\varepsilon}\vert \mathcal{M}_{\rm BN}\vert^2&=&\frac{32G_F^2a^2\alpha_{\rm QED}^3}{\pi(180)^2}\frac{\omega_0^4\omega_1\omega_2}{m_e^8}\bigg\{(1-{\bf n}_1\cdot{\bf n}_2)\Big[72\Big({\bf E}^2-({\bf E}\cdot{\bf n}_0)^2\Big)\Big({\bf B}^2-({\bf B}\cdot{\bf n}_0)^2\Big)-968\vert{\bf E}\cdot({\bf B}\times{\bf n}_0)\vert^2\Big]\non
&&\hspace{0cm}+72\,\Big[{\bf n}_1\cdot({\bf B}\times{\bf n}_0){\bf n}_2\cdot({\bf B}\times {\bf n}_0)\big({\bf E}^2-({\bf E}\cdot{\bf n}_0)^2\big)+{\bf n}_1\cdot({\bf B}\times {\bf n}_0)({\bf n}_2-{\bf n}_0)\cdot{\bf E}\,({\bf B}\times {\bf n}_0)\cdot{\bf E}+({\bf n}_1\leftrightarrow{\bf n}_2)\non
&&\hspace{1cm}+({\bf n}_1-{\bf n}_0)\cdot{\bf E}\,({\bf n}_2-{\bf n}_0)\cdot{\bf E}\,[{\bf B}^2-({\bf B}\cdot {\bf n}_0)^2]\Big]\non
&&\hspace{0cm}+392\,(1-{\bf n}_0\cdot{\bf n}_1)(1-{\bf n}_0\cdot{\bf n}_2)\Big[{\bf E}^2{\bf B}^2-({\bf E}\cdot{\bf B})^2-\vert{\bf n}_0\cdot({\bf E}\times {\bf B})\vert^2\Big]\non
&&\hspace{0cm}+784\Big[(1-{\bf n}_1\cdot{\bf n}_0){\bf E}\cdot({\bf B}\times{\bf n}_0)\,({\bf n}_0+{\bf n}_2)\cdot({\bf E}\times{\bf B})+({\bf n}_1\leftrightarrow {\bf n}_2)\Big]\non
&&\hspace{0cm}+168\Big\{({\bf E}\cdot{\bf n}_0)\,{\bf n}_0\cdot({\bf B}\times{\bf E})\Big[{\bf n}_2\cdot({\bf B}\times {\bf n}_0)(1-{\bf n}_0\cdot{\bf n}_1)+({\bf n}_1\leftrightarrow{\bf n}_2)\Big]\non
&&\hspace{1cm}+\Big[\vert{\bf B}\vert^2({\bf E}\cdot{\bf n}_0)-({\bf B}\cdot{\bf n}_0)({\bf E}\cdot{\bf B})\Big]\Big[({\bf n}_2-{\bf n}_0)\cdot{\bf E}(1-{\bf n}_0\cdot{\bf n}_1)+({\bf n}_1\leftrightarrow{\bf n}_2)\Big]\Big\}\non
&&\hspace{0cm}+672\Big[{\bf n}_1\cdot({\bf B}\times {\bf n}_0)\,({\bf n}_2-{\bf n}_0)\cdot{\bf E}+({\bf n}_1\leftrightarrow {\bf n}_2)\Big]\,{\bf E}\cdot({\bf B}\times{\bf n}_0)
\bigg\}\,.\non
\label{samp4}
\ear
\end{widetext}
The above four-vectors are defined accroding to  
\bear
&&k=\omega_0(1,{\bf n}_0)\,,\non
&&k_{\rm N}=(0,\Delta{\bf q})~~({\rm static~ photon~from~the~Coulomb~field})\,,
\non
&&p_1=p_\nu=\omega_1(1,{\bf n}_1)\,,\non
&&p_2=p_{\bar{\nu}}=\omega_2(1,{\bf n}_2)\,.
\ear
The unit vectors ${\bf n}_0$, ${\bf n}_1$, and ${\bf n}_2$ are related to the incoming photon, neutrino, and antineutrino pairs. In the next section, we consider a special configurations the magnetic field is parallel to the incoming photon momentum, i.e. ${\bf B}\parallel{\bf k}$.
\section{Parallel configuration}
Examples of specific configurations include the magnetic field and incoming photon momentum being parallel or perpendicular, i.e. ${\bf B}\parallel {\bf k}$ or ${\bf B}\perp{\bf k}$. In \cite{shai-98}, the perpendicular case was explored, resulting in the enhancement mentioned in the Introduction. However, even in the case of a pure magnetic field background and the parallel configuration, non-zero results can be obtained in the final cross section. For the remainder of this paper, we will focus only on the parallel scenario, where ${\bf B}\times {\bf n}_0=0$ and ${\bf B}^2-({\bf B}\cdot{\bf n}_0)^2=0$. In this case, (\ref{samp4}) has only one contribution.
\begin{widetext}
\bear
\half\sum_{s,s';\,\varepsilon}\vert \mathcal{M}_{ BN}\vert^2
&=&\frac{32G_F^2a^2\alpha_{\rm QED}^3}{ (180)^2\pi}\frac{\omega_0^2}{m_e^8}\bigg\{392(\omega_0\omega_1-{\bf k}\cdot{\bf p}_1)(\omega_0\omega_2-{\bf k}\cdot{\bf p}_2)\Big[{\bf E}^2{\bf B}^2-({\bf E}\cdot{\bf B})^2\Big]\bigg\}\non
&=&\frac{32\times 392\,G_F^2a^2\alpha_{\rm QED}^3}{(180)^2\pi}\frac{\omega_0^4\omega_1\omega_2}{m_e^8}\frac{Q^2}{\vert\Delta{\bf q}\vert^4}\Big(1-{\bf n}_0\cdot {\bf n}_1\Big)\Big(1-{\bf n}_0\cdot {\bf n}_2\Big)\Big(\vert\Delta{\bf q}\vert^2{\bf B}^2-(\Delta{\bf q}\cdot{\bf B})^2\Big)\non
&=&\frac{12544\,G_F^2a^2\alpha_{\rm QED}^3}{(180)^2\pi}\frac{\omega_0^4\omega_1\omega_2}{m_e^8}\frac{Q^2B^2}{\vert\Delta{\bf q}\vert^2}\Big(1-{\bf n}_0\cdot {\bf n}_1\Big)\Big(1-{\bf n}_0\cdot {\bf n}_2\Big)\Big(1-\cos^2\varphi\Big)\,,
\ear
\end{widetext}
with the momentum transfer defined 
\bear
\Delta{\bf q}={\bf k}-({\bf p}_1+{\bf p}_2)\,.
\ear
Assuming $\varphi$ is the angle between the magnetic field and the momentum transfer, the scattering angles $\theta_{1,2}$ between the incoming photon and the outgoing particles can be expressed as ${\bf n}_0\cdot({\bf n}_1,{\bf n}_2)=(\cos\theta_1,\cos\theta_2)$. Thus, we obtain:
\bear
\half\sum_{s,s';\,\varepsilon}\vert \mathcal{M}_{ BN}\vert^2
&=&\frac{12544\,G_F^2a^2\alpha_{\rm QED}^3}{(180)^2\pi}\frac{\omega_0^4\omega_1\omega_2}{m_e^4 B_{\rm crit}^2}\frac{Z^2 B^2}{\vert\Delta{\bf q}\vert^2}\non
&&\times(1-\cos\theta_1)(1-\cos\theta_2)(1-\cos^2\varphi)\,.\non
\ear

In order to establish a comparison with earlier research, we calculated our cross section by numerically evaluating the momentum integrals. We find that 
\bear
\sigma_{\rm BN}\approx 4\times 10^{-48} \alpha_{\rm QED}^3Z^2(\frac{B}{{B_{\rm crit}}})^2(\frac{\omega}{m})^6\,{\rm cm}^2.
\ear
Our result can be compared to the cross section reported by Rosenberg in 1963 \cite{ros-63}. 
Rosenberg's study pertains to the decay of an incoming photon in the presence of a Coulomb field, resulting in the production of neutrino pairs
\bear
\sigma\vert_{\gamma+Z\rightarrow\nu\bar{\nu}}\approx 1.25\times 10^{-49}\alpha_{\rm QED}^3 Z^2(\frac{\omega}{m})^6\,{\rm cm}^2.
\ear
By examining the ratio of these two cross sections, we arrive at
\bear
\frac{\sigma_{\it BN}}{\sigma\vert_{\gamma+Z\rightarrow \nu\bar{\nu}}}\approx 32 (\frac{B}{B_{\rm crit}})^2\,,
\ear
which is directly proportional to the squared dimensionless ratio of $(B/B_{\rm crit})$.

\section{Conclusions}\label{Conclusions}
In this paper, we investigated the scattering of a low energy photon in the presence of a combined field of a nucleus and an external magnetic field. This scenario was different from previous studies involving only a magnetic field which leads to an enhancment in the final cross section. The presence of the Coulomb field results in a large interaction volume in forward direction $d\sigma/d\Omega\propto 1/(\Delta{\bf q})^2$.
We calculated the differential cross section of the scattering process and found that it exhibited a peak in the forward scattering direction. The large interaction volume and the peak in the forward scattering direction observed in this study have important implications for the study of photon interactions in astrophysical environments. It suggests that the combined field of a nucleus and an external magnetic field can significantly affect the behavior of photons, which could impact the observations of astrophysical phenomena such as the energy loos of stars.
Overall, the study provides valuable insights into the complex interactions between photons and background fields, which are important for understanding the behavior of photons in astrophysical environments.


\acknowledgments 
 
R.S.~acknowledges support from Deutsche Forschungsgemeinschaft 
(DFG, German Research Foundation, grant 278162697 -- SFB 1242). 
R.Sh. was partially supported by the project “Advanced research using high intensity laser produced photons and particles” (ADONIS) (CZ.02.1.01/0.0/0.0/16 019/0000789) from the European Regional Development Fund.

\end{document}